\documentstyle[12pt]{article}
\def\bqn{\begin{equation}}
\def\nqn{\end{equation}}

\newtheorem{lemm}{Lemma}

\def\C{{\rm\kern.24em \vrule width.02em height1.4ex depth-.05ex \kern-.26em
C}}
\def\R{{\rm I\kern-.20em R}}
\def\Z{{\rm\kern.26em \vrule width.02em height0.5ex depth0ex \kern.04em
\vrule  width.02em height1.47ex depth-1ex \kern-.34em Z}}
\def\Q{{\rm\kern.24em \vrule width.02em height1.4ex depth-.05ex \kern-.26em
Q}}

\begin{document}

\begin{titlepage}
\vspace{2.5cm}
\begin{center}{\large \vspace{4.0cm} \bf Aspects of Classical
and Quantum Nambu Mechanics}\\ \vspace{1cm}
RUPAK CHATTERJEE \\
Department of Physics, State University of New York at Stony Brook \\
Stony Brook, NY 11794-3800, USA \\
e-mail: rupak@max.physics.sunysb.edu \\
\vspace{0.5cm}
and \\
\vspace{0.5cm}
LEON TAKHTAJAN \\
Department of Mathematics, State University of New York at Stony Brook \\
Stony Brook, NY 11794-3651, USA \\
e-mail: leontak@math.sunysb.edu \\
\vspace{0.5cm}
July, 1995
\vspace{3cm}

\begin{abstract}

We present recent developments in the theory of Nambu mechanics, which
include new examples of Nambu-Poisson manifolds with linear Nambu brackets
and new representations of Nambu-Heisenberg commutation relations.

\vspace{0.5cm}


\end{abstract}

\end{center}
\end{titlepage}

\section{Introduction}

Nambu mechanics is a generalization of classical Hamiltonian mechanics,
introduced by Yoichiro Nambu [1]. Recently [2], its basic principles have been
formulated in an invariant geometrical form similar to that of Hamiltonian
mechanics. In [3] new examples of classical dynamical systems were given,
which can be described by this formalism. Despite the elegance and beauty of
Nambu mechanics, it turns out be somewhat restrictive [2] with many basic
problems waiting to be solved.

This letter begins with a few new examples of Nambu-Poisson manifolds.
The main result of this section is the generalization of the linear
Poisson bracket to the Nambu bracket. In the next section we consider the
quantization of Nambu mechanics via the canonical formalism of Nambu-Heisenberg
commutation relations and present their new representations.

\section{Nambu-Poisson Manifolds}

Let $M$ denote a smooth finite dimensional manifold and $C^{\infty }(M)$
the algebra of infinitely differentiable real valued functions on $M$. Recall
that [2] $M$ is a called a Nambu-Poisson manifold if there exits a
$\R$-multi-linear map
\bqn
\{ ~,\ldots,~ \}~:~[C^{\infty }(M)]^{\otimes n} \rightarrow C^{\infty }(M)
\nqn
called a Nambu bracket of order $n$ such that
 $\forall f_1 , f_2 , \ldots , f_{2n-1} \in C^{\infty }(M)$,
\bqn
\{ f_1, \ldots ,f_n \}=(-1)^{\epsilon(\sigma)}\{ f_{\sigma(1)}, \ldots ,
f_{\sigma(n)} \},
\nqn
\bqn
\{ f_1 f_2, f_3, \ldots ,f_{n+1} \}=
f_1 \{f_2, f_3, \ldots , f_{n+1} \} +
\{ f_1, f_3, \ldots, f_{n+1} \} f_2,
\nqn
and
\begin{eqnarray}
\{ \{ f_1, \ldots , f_{n-1}, f_n \}, f_{n+1}, \ldots, f_{2n-1} \} +
\{ f_n, \{ f_1, \ldots, f_{n-1}, f_{n+1} \}, f_{n+2}, \ldots , f_{2n-1} \} \\
 +  \cdots + \{ f_n, \ldots ,f_{2n-2}, \{ f_1, \ldots , f_{n-1}, f_{2n-1} \}\}
 =  \{ f_1, \ldots , f_{n-1}, \{ f_n, \ldots , f_{2n-1} \}\}, \nonumber
\end{eqnarray}
where $\sigma \in S_n$---the symmetric group of $n$ elements---and
$\epsilon(\sigma)$ is its parity.  Equations (2) and (3) are the standard
skew-symmetry and derivation properties found for the ordinary ($n=2$) Poisson
bracket, whereas (4) is a generalization of the Jacobi identity and was called
in [2] the fundamental identity.

The dynamics on a Nambu-Poisson manifold $M$ (i.e.~a phase space) is
determined by $n-1$ so-called Nambu-Hamiltonians $H_1, \ldots, H_{n-1} \in
C^{\infty }(M)$ and is governed by the following equations of motion
\bqn
\frac {df}{dt} =  \{ f , H_1 ,\ldots,H_{n-1} \},~\forall f \in
C^{\infty }(M).
\nqn

A solution to the Nambu-Hamilton equations of motion produces an evolution
operator $U_t$ which by virtue of the fundamental identity preserves
the Nambu bracket structure on $C^{\infty }(M)$.

The Nambu bracket is geometrically realized by a Nambu tensor field $\eta \in
\wedge ^n TM$, a section of the $n$-fold exterior power $\wedge ^n TM$ of the
tangent bundle $TM$, such that
\bqn
\{ f_1,\ldots,f_n \} = \eta (df_1,\ldots,df_n),
\nqn
which in local coordinates $(x_1,\ldots,x_n)$ is given by
\bqn
 \eta = \eta_{i_1 \ldots i_n}(x) \frac{\partial}{\partial x_{i_1}} \wedge
\cdots
\wedge \frac{\partial}{\partial x_{i_n}},
\nqn
where repeated indices are assumed to be summed.

It was stated in [2] that the fundamental identity (4) is equivalent to
the following algebraic and differential constraints on the Nambu tensor
$\eta_{i_1 \ldots i_n}(x)$:
\bqn
\Sigma_{ij} + P(\Sigma)_{ij} = 0,
\nqn
for all multi-indices $i=\{i_1,\ldots,i_n \}$ and
$j=\{j_1,\ldots,j_n \}$ from the set $\{1,\ldots,N \}$,
where
\bqn
\Sigma _{ij} = \eta _{i_1 \ldots i_n} \eta _{j_1 \ldots j_n}
+ \eta _{j_n i_1 i_3 \ldots i_n } \eta _{j_1 \ldots j_{n-1} i_2}
+ \cdots + \eta _{j_n i_2 \ldots i_{n-1} i_1 } \eta _{j_1 \ldots j_{n-1} i_n}
- \eta _{j_n i_2 \ldots i_n } \eta _{j_1 \ldots j_{n-1} i_1},
\nqn
and $P$ is the permutation operator which interchanges the indices $i_1$ and
$j_1$ of $2n$-tensor $\Sigma$, and
\begin{eqnarray}
{} & {} & \sum_{l=1}^{N}  \left ( \eta_{l i_2 \ldots i_n}
\frac{\partial \eta_{j_1 \ldots j_n}}
{\partial x_l} + \eta_{j_n l i_3 \ldots i_n}\frac{\partial \eta_{j_1 \ldots
j_{n-1} i_2}}{\partial x_l} + \ldots + \eta_{j_n i_2 \ldots i_{n-1}l}
\frac{\partial\eta_{j_1 \ldots j_{n-1} i_n}}{\partial x_l}
\right) \nonumber \\
& = & \sum_{l=1}^{N}  \eta_{j_1 j_2 \ldots j_{n-1} l}
\frac{\partial \eta_{j_n i_2  \ldots i_n}}
{\partial x_l},
\end{eqnarray}
for all $i_2,\ldots,i_n,~j_1,\ldots,j_n = 1,\ldots,N$.

It was noted in [2] that the equation $\Sigma_{ij}=0$ is equivalent to the
condition that $n$-tensor $\eta$ is decomposable so that any decomposable
element in $\wedge^nV$, where $V$ is an $N$-dimensional vector space over
$\R$, endows $V$ with the structure of a Nambu-Poisson manifold. In
particular, the totally antisymmetric $n$-tensor in $\R^n$ defines a Nambu
bracket. In addition to that we have the following result.

\begin{lemm}
The completely antisymmetric constant $(n-1)$-tensor
${\eta}_{i_1 \ldots i_{n-1}}$, where
$i_1,\ldots,i_{n-1}=1,\ldots, n$ is a Nambu tensor.
\end{lemm}
{\bf Proof}. Note that $\eta$ is a $n-1$-tensor in $n$-dimensional
space and not in $n-1$-dimensions. It is sufficient to prove that
$\eta$ is decomposable. For the case $n=4$ one has
\begin{eqnarray}
{} & {} & {\eta}_{ijk}~ \frac{\partial}{\partial x_i}
\wedge \frac{\partial}{\partial x_j} \wedge
\frac{\partial}{\partial x_k}
=   \left( \frac{\partial}{\partial x_1}
\wedge \frac{\partial}{\partial x_2} \wedge
\frac{\partial}{\partial x_3} \right)
+  \left( \frac{\partial}{\partial x_1}
\wedge \frac{\partial}{\partial x_2} \wedge
\frac{\partial}{\partial x_4} \right) \nonumber \\
{} & {} & + \left(  \frac{\partial}{\partial x_1}
\wedge \frac{\partial}{\partial x_3} \wedge
\frac{\partial}{\partial x_4} \right)
+  \left( \frac{\partial}{\partial x_2}
\wedge \frac{\partial}{\partial x_3} \wedge
\frac{\partial}{\partial x_4}  \right)  \\
{} & {} & = \frac{1}{4} \left[ \left( \frac{\partial}{\partial x_1}
+\frac{\partial}{\partial x_2}+\frac{\partial}{\partial x_3}
+\frac{\partial}{\partial x_4} \right)
{}~\wedge ~ \left( \frac{\partial}{\partial x_1}
-\frac{\partial}{\partial x_2}
-\frac{\partial}{\partial x_3}
+\frac{\partial}{\partial x_4}\right) \right. \nonumber \\
{} & {} & \wedge \left. \left(\frac{\partial}{\partial x_1}
+\frac{\partial}{\partial x_2}
-\frac{\partial}{\partial x_3}
-\frac{\partial}{\partial x_4} \right) \right]. \nonumber
\end{eqnarray}

For general $n$ a similar elementary exterior algebra proof can be given.
One can also show that completely antisymmetric constant $n$-tensor
${\eta}_{i_1 \ldots i_{n-1}a}$ in $\R^N$, where $i_1,\ldots,i_{n-1}=1,\ldots,n$
and $a=n+1,n+2,\ldots,N$ is decomposable and, therefore, is a Nambu tensor.
This gives a triple Nambu bracket on the vector space $\R^N$ for any $N \geq
3$.

As an example, one can use this bracket to describe the integrable system of
two vortices in an incompressible ideal fluid with the Hamiltonian given by
\bqn
H= \ln{[ (q_1 - q_2 )^2 + (p_1 - p_2 )^2]}.
\nqn
This system has the integrals of motion
\begin{eqnarray}
I_1 & = & q_1 + q_2 \nonumber \\
I_2 & = & p_1 + p_2 \\
I_3 & = & p_1 ^2 + p_2 ^2 + q_1 ^2 + q_2 ^2 \nonumber
\end{eqnarray}
and was realized as a Nambu four bracket in [3]. It can be also represented via
a Nambu triple bracket as follows:
\bqn
\{f,I_1,H \} = (1/2)\eta_{ijk}\frac{\partial f}{\partial x_i}
\frac{\partial I_1}{\partial x_j}\frac{\partial H}{\partial x_k},
\nqn
where $i,j,k=1,2,3,4$ and $x_1=p_1,~x_2=p_2,~x_3=q_1,~x_4=q_2$.
One can also represent it as
\bqn
\{f,I_2,H \}  = (-1/2)\eta_{ijk} \frac{\partial f}{\partial x_i}
\frac{\partial I_2}{\partial x_j} \frac{\partial H}{\partial x_k}.
\nqn

So far we have been dealing with constant Nambu tensors. Next we shall
consider Nambu tensors linear in their coordinates. Namely, let $V$ be a
$n+1$-dimensional vector space over $\R$ with coordinates $x_1,
\ldots, x_{n+1}$ and define the following $n$-tensor
\bqn
\eta _{i_1 \ldots i_n} (x)=\epsilon ^
{i_{n+1}}_{i_1 \ldots  i_n}x_{i_{n+1}},
\nqn
where $\epsilon^{i_{n+1}}_{i_1 \ldots i_n}=\epsilon_{i_1 \ldots i_{n+1}}$ is
a completely antisymmetric $n+1$-tensor.

\begin{lemm} The tensor $\eta _{i_1 \ldots i_n}(x)$ is a Nambu tensor.
\end{lemm}
{\bf Proof}. First consider the $n$=3 case:
\bqn
\eta _{ijk}(x) = {\epsilon^l}_{ijk}x_l,
\nqn
where $i,j,k,l=1,2,3,4$. We must show that the tensor $\eta_{ijk}(x)$
satisfies both the algebraic and differential conditions (8)-(10). First
consider the differential constraint (10), which reads
\bqn
(\epsilon_{ijkl}\epsilon_{mnoi}+\epsilon_{oikl}\epsilon_{mnji}+
\epsilon_{ojil}\epsilon_{mnki}-\epsilon_{mnil}\epsilon_{ojki})x_l=0.
\nqn
In order to prove it, write down the decomposability condition $\Sigma_{ij}=0$,
$i,j=1,2,3,4$ for the $4$-tensor $\epsilon_{ijkl}$
\bqn
\epsilon_{ijkl}\epsilon_{mnop}+\epsilon_{oikl}\epsilon_{mnjp}-\epsilon_{oijl}
\epsilon_{mnkp}+\epsilon_{ojki}\epsilon_{mnlp}-\epsilon_{ojkl}\epsilon_{mnip}=0.
\nqn
Setting $p=i$ in the above expression gives
\bqn
\epsilon_{ijkl}\epsilon_{mnoi}+\epsilon_{oikl}\epsilon_{mnji}+\epsilon_{ojil}
\epsilon_{mnki}-\epsilon_{ojki}\epsilon_{mnil}=0,
\nqn
which is equivalent to the equation (18). Now, the algebraic condition (8)
for $\eta _{ijk}(x)$ reads:
\begin{eqnarray}
(\epsilon_{i_1i_2i_3l}\epsilon_{j_1j_2j_3k}+\epsilon_{ki_1i_3l}\epsilon_{j_1j_2
j_3i_2}+\epsilon_{ki_2i_1l}\epsilon_{j_1j_2j_3i_3}+ \nonumber \\
\epsilon_{j_1i_2i_3l}\epsilon_{i_1j_2j_3k}+\epsilon_{kj_1i_3l}\epsilon_{i_1j_2
j_3i_2}+\epsilon_{ki_2j_1l}\epsilon_{i_1j_2j_3i_3})x_lx_{j_3}=0.
\end{eqnarray}
To prove it, write down the equation $\Sigma_{ij} + P(\Sigma_{ij})=0$
for $\epsilon_{ijkl}$ and contract it with two $x$'s as follows:
\begin{eqnarray}
(\epsilon_{i_1i_2i_3l}\epsilon_{j_1j_2j_3k}+\epsilon_{ki_1i_3l}\epsilon_{j_1j_2
j_3i_2}+\epsilon_{ki_2i_1l}\epsilon_{j_1j_2j_3i_3}+\epsilon_{ki_2i_3i_1}
\epsilon_{j_1j_2j_3l}+ \nonumber \\
\epsilon_{j_1i_2i_3l}\epsilon_{i_1j_2j_3k}+\epsilon_{kj_1i_3l}\epsilon_{i_1j_2
j_3i_2}+\epsilon_{ki_2j_1l}\epsilon_{i_1j_2j_3i_3}+\epsilon_{ki_2i_3j_1}
\epsilon_{i_1j_2j_3l})x_l x_{j_3}=0.
\end{eqnarray}
The fourth and eighth terms in the above equation vanish since
$\epsilon_{j_1j_2j_3l}x_lx_{j_3} =0$ and thus equation (22) is equivalent
to equation (21). This completes the proof for $n=3$; the proof for $n>3$
is similar.
\vskip 0.5cm

{\bf Remark 1} This construction generalizes the standard linear Poisson
bracket on a vector space $V$, given by the Poisson tensor $\eta _{ij}(x)=
c^{k}_{ij} x_k$. The Jacobi identity for this bracket is equivalent to the
Jacobi identity for the structure constants $c^{k}_{ij}$ so that the
dual space $g=V^*$ has a Lie algebra structure. This classical construction
goes back to Sophus Lie and plays a fundamental role in representation
theory. In [2] and [4] a generalization of Lie algebras to the case of higher
order operations, called Nambu-Lie gebras, were introduced. The example of a
linear Nambu bracket (16) provides the dual space $V^*$ with the Nambu-Lie
structure.

{\bf Remark 2} Note that algebraic constraint on a Nambu tensor
$\eta_{ijk}=c _{ijk}$ according to [2, Remark 1] contains the Jacobi identity
for $c_{ijk}$ interpreted as structure constants. Namely, writing down the
algebraic constraint for Nambu tensor $c_{ijk}$:
\begin{eqnarray}
c_{i_1i_2i_3}c_{j_1j_2j_3}+c_{j_3i_1i_3}c_{j_1j_2i_2}+c_{j_3i_2i_1}
c_{j_1j_2i_3}+  \nonumber \\
c_{j_1i_2i_3}c_{i_1j_2j_3}+c_{j_3j_1i_3}c_{i_1j_2i_2}+c_{j_3i_2j_1}
c_{i_1j_2i_3}=0,
\end{eqnarray}
and setting $j_2 = i_1$ gives
\bqn
c_{i_1i_2i_3}c_{j_1i_1j_3}+c_{j_3i_1i_3}c_{j_1i_1i_2}+c_{j_3i_2i_1}
c_{j_1i_1i_3}=0,
\nqn
which is the Jacobi identity for the totally skew-symmetric structure
constants $c_{ijk}$. This opens a possibility of using the structure
constants of simple Lie algebras as Nambu tensors. Unfortunately, except for
the structure constants of $sl(2,\C)$, they do not satisfy the fundamental
identity and, therefore, can not serve as Nambu tensors. Indeed, it follows
from the Cartan classification that either $sl(3,\C)$ or $sl(2,\C) \oplus
sl(2,\C)$ are subalgebras of all simple Lie algebras. However, it can be shown
directly that the structure constants for these Lie algebras do not satisfy
the fundamental identity.

So far, all examples of Nambu tensors turn out to be decomposable. This leads
us to a conjecture that all Nambu tensors are decomposable. In other words,
equation $\Sigma_{ij}+P(\Sigma_{ij})=0$ for all multi-indices $i$ and $j$
should imply that $\Sigma_{ij}=0$.
\vskip 1.0cm

\section{Representations of Nambu-Heisenberg Commutation Relations}

There exist different approaches towards a quantization of Nambu mechanics,
such as deformation quantization in the spirit of [5] and Feynman path
integral approach, based on the action principle for Nambu mechanics
[2]. Here we will comment only on the method of canonical quantization,
which is based on the Heisenberg commutation relations
\bqn
[a,a^{\dagger}]= aa^{\dagger}-a^{\dagger}a = I,
\nqn
and its generalization to higher order algebraic structures proposed by Nambu
[1]:
\bqn
[A_1,\ldots,A_n] =^{def} \sum_{\sigma \in S_n}(-1)^{\epsilon(\sigma)}
A_{\sigma(1)}A_{\sigma(2)}\cdots A_{\sigma(n)} =cI,
\nqn
where $I$ is the unit operator and $c$ is a constant. In [2] it was called
the Nambu-Heisenberg commutation relation. In particular, for $n=3$ we have
\bqn
[A_1,A_2,A_3] = A_1 A_2 A_3 - A_1 A_3 A_2 + A_3 A_1 A_2 -
A_3 A_2 A_1 + A_2 A_3 A_1 - A_2 A_1 A_3.
\nqn

Denote by $\zeta$ the primitive $n^{th}$ root of unity, $\zeta^n=1$, and by
$\Q[\zeta]$ the corresponding cyclotomic field: an algebraic extension of the
field of rational numbers $\Q$ by $\zeta$. When $n$ is a prime number, the
minimal polynomial for $\zeta$ has the form $1+\zeta+\zeta^2+\cdots+\zeta^{n-1}
=0$. Let $\Z[\zeta]$ be a ring of algebraic integers in $\Q[q]$, i.e.~
$$\Z[\zeta]=\{\omega=m_1+m_2\zeta+m_3\zeta^2+\cdots+m_{n-2}\zeta^{n-2}~|~m_1,
\ldots,m_{n-2} \in \Z \},$$
and let ${\cal H}_n$ will be a vector space over $\C$ with a basis
$\{|\omega>\}$, parametrized by the elements $\omega \in \Z[\zeta]$.

It was shown in [2] that the $n=3$ Nambu-Heisenberg commutation relation
admits the following representation:
\vskip 0.5cm
\begin{eqnarray}
A_1|\omega> & = & (\omega+\zeta+1)|\omega+1>, \nonumber \\
A_2|\omega> & = & (\omega+\zeta)|\omega+\zeta>, \\
A_3|\omega> & = & \omega|\omega+\zeta^2>, \nonumber
\end{eqnarray}
\noindent where $|\omega> \in {\cal H}_3 $ and $c=\zeta^2(\zeta+1)$.

In [2] it was also mentioned that the Nambu-Heisenberg commutation relations
for
general $n$ admit a natural representation in the vector space ${\cal H}_n$.
Here we present the following result, which can be proved directly (using
a symbolic calculations package).

\begin{lemm} The Nambu-Heisenberg commutation relations for $n=5$ and $n=7$
admit the following representations.
\vskip 0.5cm

\noindent 1) $[A_1,\ldots,A_5]=(-2\zeta^4+3\zeta^2+3\zeta-2-2\zeta^{-2})I$
\begin{eqnarray}
A_1|\omega> & = & (\omega+\zeta^3+\zeta^2+\zeta+1)|\omega+1>,
\nonumber \\
A_2|\omega> & = & (\omega+\zeta^3+\zeta^2+\zeta)|\omega+\zeta>,
\nonumber \\
A_3|\omega> & = & (\omega+\zeta^3+\zeta^2)|\omega+\zeta^2>, \\
A_4|\omega> & = & (\omega+\zeta^3)|\omega+\zeta^3> \nonumber \\
A_5|\omega> & = & \omega ^2|\omega+\zeta^4>, ~~|\omega> \in {\cal H}_5.
\nonumber
\end{eqnarray}
\vskip 0.5cm
\noindent 2) $[A_1,\ldots,A_7]=7(3+3\zeta^{-1}+\zeta^{-2}-3\zeta^2+\zeta^{-3}-
3\zeta^3)I$
\begin{eqnarray}
A_1|\omega> & = & (\omega+\zeta^5+\zeta^4+\zeta^3+\zeta^2+\zeta+1)|\omega+1>
\nonumber \\
A_2|\omega> & = & (\omega+\zeta^5+\zeta^2+\zeta^3+\zeta^2+\zeta)|\omega+\zeta>
\nonumber \\
A_3|\omega> & = & (\omega+\zeta^5+\zeta^4+\zeta^3+\zeta^2)|\omega+\zeta^2>
\nonumber \\
A_4|\omega> & = & (\omega+\zeta^5+\zeta^4+\zeta^3)|\omega+\zeta^3> \\
A_5|\omega> & = & (\omega^2+\zeta^5+\zeta^4)|\omega+\zeta^4> \nonumber \\
A_6|\omega> & = & (\omega+\zeta^5)|\omega+\zeta^5> \nonumber \\
A_7|\omega> & = & \omega^3|\omega+\zeta^6>,~~~|\omega> \in {\cal H}_7.
\nonumber
\end{eqnarray}
\end{lemm}
{\bf Remark 3} Although we are absolutely certain that there exists a natural
representation for any $n$, we were unable to construct it explicitly. We only
mention that somehow composite and prime $n$'s differ dramatically in this
respect (which shows in representations for $n=4$ and $n=6$, which we do not
present here) and it is rather difficult to present explicit constructions for
prime $n$.
\vskip 0.5cm

One possibility of realizing these representations was suggested to us by
Igor Frenkel, who proposed that operators $A_i$ may be interpreted as
special difference operators acting on a linear span of a root lattice for
simple Lie algebras. Namely, for $n=3$ consider the root system $\Phi$ of
the Lie algebra ($sl(3,\C)$). It consists of six roots $\pm \alpha=\pm(1,0),
{}~\pm\beta=\pm(-1,\sqrt{3})/2$, and $\pm\theta=\pm(\alpha+\beta)$. The simple
roots can be chosen to be $\alpha $ and $\beta$ and the Weyl group of $\Phi$,
which is isomorphic to $S_3$, is given by $W(\Phi)=\{1,~r_\alpha,~r_\beta,~
r_\alpha r_\beta,~r_\beta r_\alpha,~r_\alpha r_\beta~r_\alpha\}$, where
$r_\alpha$ and $r_\beta$ stand for corresponding reflections. The root lattice
$\Lambda_r$ of $sl(3,\C)$ is isomorphic to $\Z[\zeta]$ with $\zeta^3=1$ via
$\alpha \rightarrow 1,~\beta \rightarrow (-1+\sqrt{-3})/2 =\zeta$, so that the
vector space ${\cal H}_3$ can be parametrized by the simple roots of $sl(3,\C)$
$$|\omega>=|m_1\alpha+m_2\beta>~m_1,m_2 \in \Z.$$
For any $\sigma \in \Lambda_r$ define a shift operator $A_\sigma$ in
${\cal H}_3$ by
\bqn
A_\sigma|\omega>=[r_\alpha r_\beta (\sigma)+\omega]|\omega+\sigma>.
\nqn
For the case $n=3$ we can write down the representation of the Nambu-Heisenberg
commutation relations in terms of these operators as
$$[A_\alpha,~A_\beta,~A_{-\theta}] =9I,$$
or, in the previous notation,
\begin{eqnarray*}
A_1|\omega> & = & (\omega+\zeta)|\omega+1>, \\
A_2|\omega> & = & (\omega-\zeta-1)|\omega+\zeta>, \\
A_3|\omega> & = & (\omega+1)|\omega+\zeta^2>,
\end{eqnarray*}
since $r_\alpha r_\beta (\alpha)=\beta \rightarrow \zeta,~ r_\alpha r_\beta
(\beta)=-\theta \rightarrow -\zeta-1,~r_\alpha r_\beta ~(-\theta)=\alpha
\rightarrow 1$. We wonder if similar representations exist for difference
operators related with other root systems, namely with that of $sl(n,\C)$.
\vspace{0.5cm}

\noindent{\bf Acknowledgements} \\
We are grateful to Professor I.~Frenkel for helpful discussions. This work was
supported in part by the U.S.~Department of Energy grant DE-F602-888ER40388
(R.C.) and by the NSF grant DMS-92-04092 (L.T.).

\section{References}

1. Nambu, Y., {\em Phys.~Rev.~D. } {\bf 7}, 2405 (1973).\\
2. Takhtajan, L.A., {\em Comm.~Math.~Phys. } {\bf 160}, 295 (1994). \\
3. Chatterjee, Rupak, {\em Lett.~Math.~Phys.} (1995). \\
4. Takhtajan, L.A., {\em St. Petersburg Math. J.} {\bf 6}, 429 (1994). \\
5. Bayen, F., Flato, M., Fronsdal, C., Lichnerowicz, A., Sternheimer, D.,
   {\em Ann.~Phys.} {\bf 110}, 67 (1978).
\end{document}